\documentclass[twocolumn,prb,citeautoscript,superscriptaddress]{revtex4}
\usepackage{amssymb}
\usepackage{amsmath}
\usepackage{graphicx}
\usepackage{bm}
\usepackage{epstopdf}

\begin{document}
\title{Strongly interacting confined quantum systems in one dimension}
\author{A.~G. Volosniev}
\author{D.~V. Fedorov}
\author{A.~S. Jensen}
\affiliation{Department of Physics and Astronomy, Aarhus University, Ny Munkegade 120, DK-8000 Aarhus C, Denmark}
\author{M. Valiente}
\affiliation{SUPA, Institute for Photonics and Quantum Sciences, Heriot-Watt University, Edinburgh EH14 4AS, United Kingdom}
\author{N.~T. Zinner}
\affiliation{Department of Physics and Astronomy, Aarhus University, DK-8000 Aarhus C, Denmark}

\begin{abstract}
In one dimension, the study of
magnetism dates back to the dawn of quantum mechanics 
when Bethe solved the famous Heisenberg model
that describes quantum behaviour in magnetic systems.
In the last decade, one-dimensional systems have
become a forefront area of research
driven by the realization of the Tonks-
Girardeau gas using cold atomic gases. Here we
prove that one-dimensional fermionic and bosonic
systems with strong short-range interactions are
solvable in arbitrary confining geometries by in-
troducing a new energy-functional technique and
obtaining the full spectrum of energies and eigen-
states. As a first application, we calculate spatial
correlations and show how both ferro- and anti-
ferromagnetic states are present already for small
system sizes that are prepared and studied in
current experiments. Our work demonstrates
the enormous potential for quantum manipula-
tion of magnetic correlations at the microscopic
scale.
\end{abstract}

\maketitle

Strongly interacting quantum particles are ubiquitous in nature and play a vital 
role in superfluidity, superconductivity, and magnetism, and
low-dimensional magnetic systems have great
potential to deliver key insights into fundamental properties of the materials 
used in modern technology.\cite{deshpande2010} 
The study of magnetism in one dimension (1D) goes back to the dawn of 
quantum mechanics when Bethe solved the Heisenberg model by introducing
the famous Bethe ansatz method.\cite{bethe1931} During the 1960s 
the technique was used to solve several paradigmatic models
including 
the Lieb-Liniger model of repulsive bosons \cite{lieb1963}, 
impurity problems with fermions by McGuire \cite{mcguire1965}, 
and Yang's solutions of homogeneous two-component Fermi systems \cite{yang1967}
that led to Lieb and Wu's solution of the 1D Hubbard
model in 1968 \cite{lieb1968}.
Common to the theoretical models is that 
periodic or open boundaries are essential ingredients, and the 
Bethe ansatz cannot be applied for
general confinement which is nevertheless the reality in many
state-of-the-art setups that have for instance been used to realize 
the Tonks-Girardeau \cite{tonks1936,girardeau1960} gas 
using cold bosonic atoms \cite{paredes2004,kino2004,haller2009}.

We present a new functional method that is specifically designed to overcome 
these problems and include the external confining geometry exactly 
in the strongly interacting regime to linear order in the inverse
interaction strength.
Furthermore, it can be used to interpolate from few to 
mesoscopic particle numbers and address how
spatial correlations emerge and evolve.\cite{serwane2011,gerhard2012,greif2013,wenz2013}
Our basic example below is a 
four-body system of two spin up and two spin down fermions
in an external trap that is solved exactly for the first
time here. Direct access to the wave function allows us to 
see ferro- and antiferromagnetic correlations in the 
eigenstates and give exact probabilities for these 
configurations. This is an extremely important feature
of this new approach as 
multi-component systems become highly degenerate
for strong interactions, and thus the energy itself yields
little information about the system.
Furthermore, we solve exactly the impurity or polaron problem of one spin 
down interacting strongly with a number of spin up particles in a harmonic
trap, a setup that has been realized experimentally \cite{wenz2013}. 
This allows us to compare different confining potentials
and show that correlations are strongly influenced by
the geometry. This has ramifications on density 
functional approaches for strong interactions, and 
our scheme can provide invaluable benchmarks of procedures where 
Bethe ansatz solutions obtained with periodic boundary
conditions are supplemented by the local density approximation. 
For finite particle numbers, periodic boundary conditions is a 
strong assumption whose justification can now be addressed using
exact solutions. Ultimately, our method could realize the goal 
of answering the key question: How many particles does it take
to make a strongly interacting many-body system?

\vspace{1em}
\noindent {\bf Results}\\
{\bf Derivation of the energy functional.}
The general system we consider has $N$ particles of mass $m$ with coordinates $x_1,\ldots,x_N$ and
is described by the Hamiltonian
\begin{align}\label{hamil}
H=\sum_i \left[p_{i}^{2}/2m+V(x_i)\right]+g\sum_{i>j}\delta(x_i-x_j),
\end{align}
where $p_i$ is the momentum operator of particle $i$ and $V(x_i)$ is an external
confining potential. We assume a short-range 
two-body interaction that we model by a Dirac delta function of strength $g$.
In the following we are interested in the strongly interacting limit where
$g\to\infty$ (or $1/g\to 0$). For simplicity, our focus will be the repulsive case ($g>0$), although 
our results can be extended linearly to the attractive side ($g<0$) of
$1/g=0$. The deeply bound states for $g\to -\infty$ \cite{lindgren2013,gharashi2013} 
are irrelevant for our arguments and will not be addressed here.
The external potential produces an energy scale, $\epsilon$, and 
a length scale, $l$, in which we will express all other quantities. In the 
examples below we will consider a double-well potential, $V(x)=m\omega^2(|x|-b)^2/2$, 
with barrier parameter, $b$, and a hard wall potential of length $L$ which vanishes
for $0<x<L$ and has infinite strength for $x<0$ and $x>L$.
A double-well potential has recently been
realized experimentally \cite{selim2013,greif2013}. 
For $b=0$, the double-well reduces to the harmonic oscillator
potential from which we adopt our units of length, $l=\sqrt{\hbar/m\omega}$, and
energy, $\epsilon=\hbar\omega$. Here $\omega$ is the oscillator angular frequency
and $\hbar=h/2\pi$ is the reduced Planck's constant. For the hard wall, we 
have $l=L$ and $\epsilon=\hbar^2/2mL^2$.
Henceforth, we will 
meaure lengths, energies, and $g$ in these units. Our
focus will be on fermions with two 
internal spin states, up and down. Systems with more 
than two internal states can be addressed in similar 
fashion (see Methods).

A general 
eigenstate wave function has the form $\Psi(x_1,\ldots,x_N)$. 
For simplicity we omit the coordinates from now on.
The zero-range interaction implies that $\Psi$ obeys the 
boundary conditions
\begin{equation}\label{bound}
\frac{1}{2g}\left[ \left(\frac{\partial\Psi}{\partial x_i}-\frac{\partial\Psi}{\partial x_j}\right)_{+}-
\left(\frac{\partial\Psi}{\partial x_i}-\frac{\partial\Psi}{\partial x_j}\right)_{-}\right]=\Psi(x_i=x_j),
\end{equation}
where the $\pm$ subscripts indicate the limits $x_i-x_j\to 0^\pm$, i.e. they are derivatives 
from each side of the point $x_i=x_j$. In the limit where $1/g\to 0$, 
the boundary conditions and the Pauli principle imply that $\Psi$ must
vanish whenever $x_i=x_j$ for any $i$ and $j$. Such functions can 
be constructed from the eigenstates of the single-particle Hamiltonian (the first term in (\ref{hamil})) 
by taking the antisymmetrized product of $N$ states.
This state we denote $\Psi_\textrm{A}$. Its energy, $E_\textrm{A}$, is a sum of the occupied single-particle
energies. However, the boundary conditions allow us to write a more general 
state on the basis of $\Psi_\textrm{A}$ \cite{girardeau1960,ogata1990,deuret2008,guan2009}, 
\begin{align}\label{psi}
\Psi=\sum_{k=1}^{N!}a_k\theta(x_{P_k(1)},\ldots,x_{P_k(N)})\Psi_\textrm{A}(x_1,\ldots,x_N),
\end{align}
where we sum over the $N!$ permutations, $P_k$, of the $N$ coordinates, and 
$\theta(x_1,\ldots,x_N)=1$ when $x_1<x_2<\ldots<x_N$ and zero otherwise. 
Fortunately, symmetries reduce the number of independent $a_k$ coefficients. 
For the present case of two-component fermions, the Pauli principle 
dictates that there are only $M=N!/(N_\uparrow!N_\downarrow!)$ degrees of
freedom. This is the number of degenerate states at $1/g=0$ which 
shows that the functions in (\ref{psi}) constitute a basis.
The basic idea is now very simple. To linear order in $1/g$ we can 
write $E=E_\textrm{A}-K/g$, where $K=K(a_1,\ldots,a_M)$ is a functional of the 
$a_k$ coefficients, and is independent of $g$ by the 
Hellmann-Feynman theorem.
One can now prove that 
$K$ has the simple form
(see Methods)
\begin{equation}\label{functional}
K=\frac{\sum_{k \neq p}\alpha_{k,p}(a_k-a_p)^2}{\sum_k a_{k}^{2}},
\end{equation}
where $k$ and $p$ run from 1 to $M$, and 
$\alpha_{k,p}$ are matrix elements that depend
only on the single-particle potential, $V(x)$.
The eigenfunctions and eigenenergies to order $1/g$
can now be obtained by variation of $K$ with respect
to $a_k$ and diagonalizing the resulting matrix.
$K$ is equivalent to Tan's contact 
parameter \cite{tan2008}
in 1D \cite{barth2011} and we compute it exactly for
$1/g\to 0$. Furthermore, the derivation can be 
easily extended to multi-component bosons, fermions
or mixtures, and provides an effective Hamiltonian 
that can be used to study perturbations in the 
strongly interacting limit (see Methods for details).

\begin{figure*}[t!]
\centering
\includegraphics[scale=0.6]{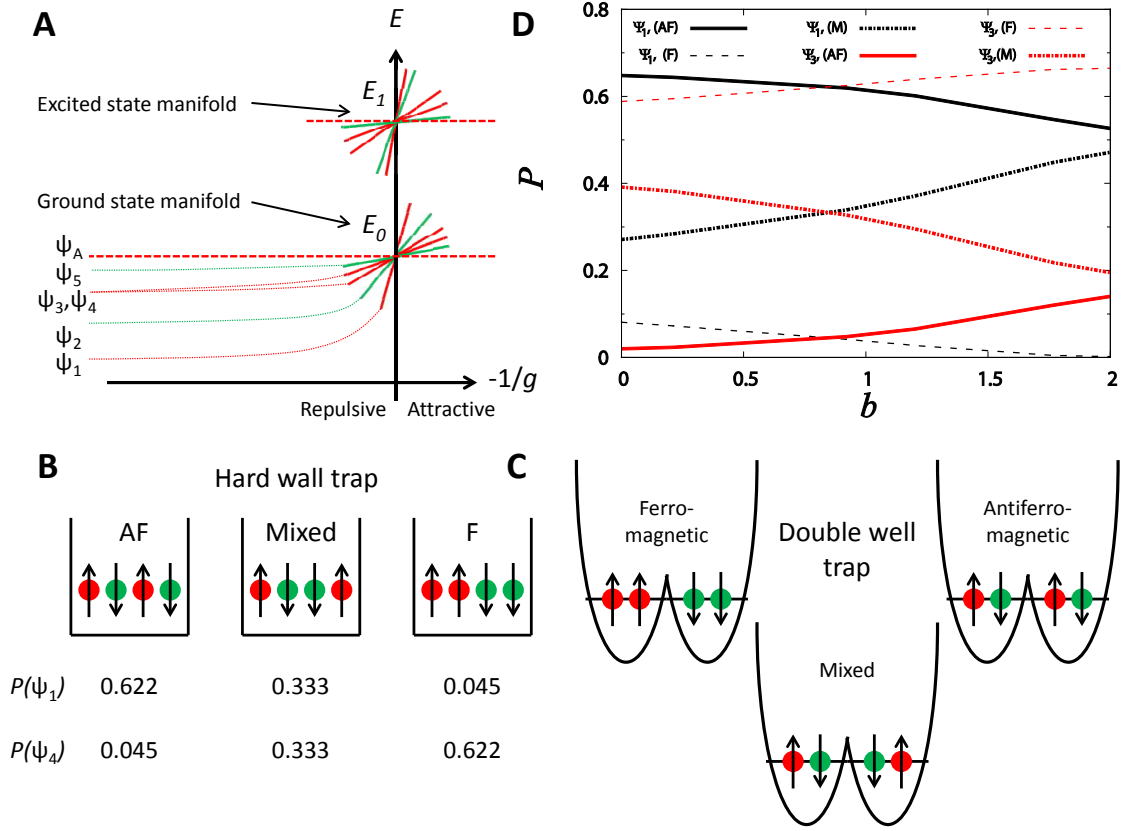}
\caption{{\bf Four-body system with two spin up and two spin down particles.} {\bf A}
Schematic spectrum of eigenstates showing the ground state and an excited state manifold. The 
slopes at infinite $g$ will generally be different around $E_0$ and $E_1$. For the ground state we
indicate the adiabatic connection between the strongly and weakly interaction regime for a harmonic trap ($b=0$). The 
red lines are positive while green lines are negative parity states. The structure around $1/g=0$ is the same
for both double well and hard wall traps, although the slopes are different.
{\bf B} Hard wall trap configurations and their probabilities for $\Psi_1$ and $\Psi_4$
which have opposite antiferromagnetic (AF) and ferromagnetic (F) contributions. The probabilities include both
the configurations shown and their parity inverted partners which are equal. 
{\bf C} Same as {\bf B} for the double well trap. {\bf D} The probability to find the three configurations in {\bf C} 
as function of the barrier parameter, $b$, for the ground state, $\Psi_1$, and an excited state, $\Psi_3$.}
\label{2plus1}
\end{figure*}

Remarkably, in the 
strongly interacting regime, the effective Hamiltonian
can always be written as a spin model. In the 
important case of $N$ spin 1/2 fermions or two-component 
bosons governed by the Hamiltonian in (\ref{hamil}), it is a 
Heisenberg model of the form 
$\sum_{i=1}^{N-1} J_i {\bm S_i}\cdot{\bm S_{i+1}}$,\cite{deu2013}
which is a favourite starting point for research 
into quantum magnetism.\cite{auerbach1998} Here
${\bm S_i}$ is the spin operator of particle $i$.
It has been shown \cite{ogata1990} that for a 
half-filled Hubbard model, $J_i=J$ is constant. This is 
obtained by using the Bethe ansatz and by the same method
one can also prove that $J_i=J$ for particles in hard wall
(box) confinement.\cite{oelkers2006} Our 
approach generalizes these important results and
not only do we find that $J_i$ depends on the external 
confinement, but also provide a procedure for computing
these nearest-neighbour interaction coefficients 
exactly. We may therefore use the 
external confinement to tailor the $J_i$ coefficients
into desirable spin models and thus manipulate 
static and dynamic quantum magnetic correlations.\cite{vol2014}
Note also that this is not only true for the 
ground state manifold, but also for higher manifolds
as illustrated in Fig.~\ref{2plus1}A. Each manifold 
will have its own unique set of $J_i$ coefficients 
that we can compute exactly. In the language of the 
Hubbard model, one can think of higher manifolds as
belonging to higher bands.

\vspace{1em}
\noindent {\bf Four-body systems.}
\noindent A central example is the hitherto unsolved four-body problem since
it illustrates the method and the magnetic correlation physics
it allows us to address.
We take $N_\uparrow=N_\downarrow=2$ with
an $M=4!/(2!)^2=6$-fold degeneracy at $1/g=0$ as
shown in Fig.~\ref{2plus1}A. In general, the spectrum around $1/g\to 0$ has the 
form of a ladder of manifolds each of which contains an $M$-fold 'fan' of states
as illustrated in Fig.~\ref{2plus1}A.
For the ground state manifold we 
also show the adiabatic connection of states from weak to 
strong coupling for the case of a harmonic trap ($b=0$) where 
the third and fourth excited states
are initially degenerate at $g=0$.
The parity invariance of the double well and hard wall potentials means that 
the three types of spatially correlated states shown in Fig.~\ref{2plus1}B and \ref{2plus1}C,
ferromagnetic, antiferromagnetic, and mixed, completely
specify all solutions at $1/g=0$.
Fig.~\ref{2plus1}C shows the configuration probabilities
for $\Psi_1$ and the state $\Psi_4$ for the hard wall which turns
out to have exact opposite ferro- and antiferromagnetic probabilities compared
to the ground state. In Fig.~\ref{2plus1}D we show 
the double well probabilities as functions of $b$ for 
$\Psi_1$ and $\Psi_3$, again picked as examples 
because of their significantly different correlations.
In both cases 
we find a ground state which is dominantly spatially
antiferromagnetic, and perhaps more remarkably we find
excited states that are dominantly spatially ferromagnetic. 
Preparing different
states at $g=0$ and then tuning to $1/g=0$ \cite{gerhard2012,wenz2013} 
would thus produce
completely different correlation patterns. Note that if one
considers two-component bosons
instead of fermions the results are very different (see Methods).

One can understand
intuitively what is going on by looking at (\ref{functional}).
The antiferromagnetic configuration is favored since $(a_i-a_j)^2$
is large for $a_i$ and $a_j$ differing in both sign and magnitude. 
The functional approach presented here thus provides a very 
precise mathematical insight into the preference for domain 
walls of opposite spin in the strongly interacting regime, and
provides a spatial explanation of antiferromagnetism in 
repulsively interacting 1D systems. Moreover, our results also
demonstrate the potential for manipulating correlations
by state preparation and trap shape modulation. A step in 
this direction was recently reported using anisotropic optical 
lattices \cite{greif2013}. 
Most often one discusses ferromagnetism induced
by symmetry breaking. We clearly have the presence
of a degenerate manifold of states to induce
such breaking and a small spin gradient is enough to 
drive the system into a purely ferromagnetic state.
However, our direct access to the exact wave function 
demonstrates the presence of intrinsic magnetic
correlations even without breaking the spin symmetry.
More generally, our method can be used to study the
correlations that drive quantum phase
transitions in larger systems using exact wave functions.

\begin{figure*}
\centering
\includegraphics[scale=0.6]{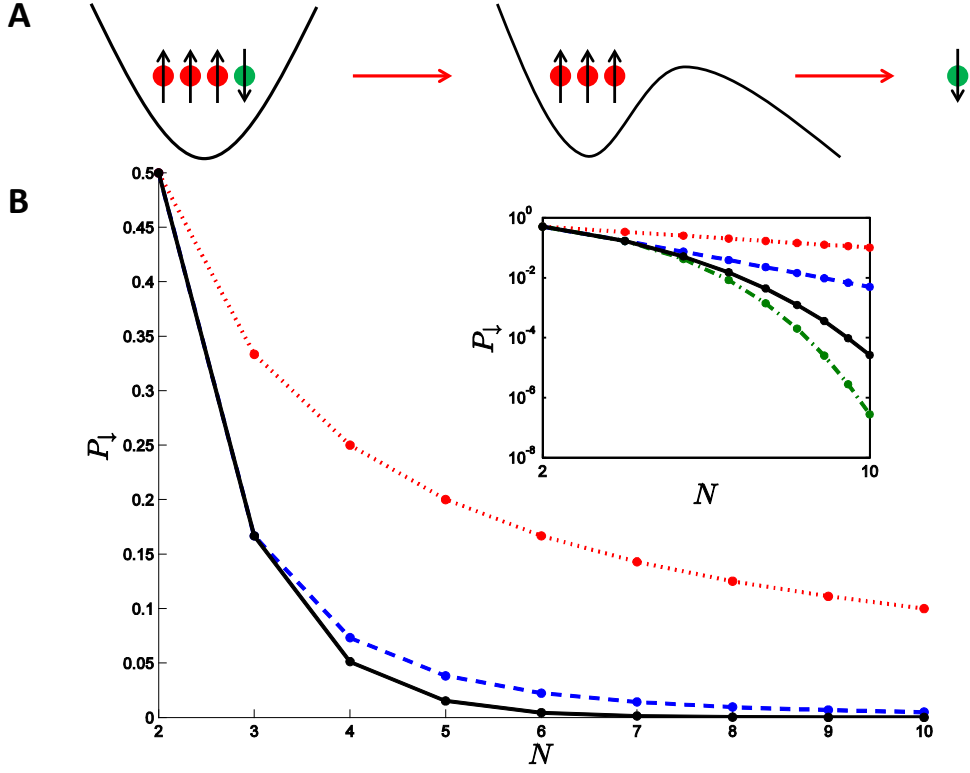}
\caption{{\bf Tunneling of a strongly interacting impurity in a Fermi sea.}
{\bf A} Illustration of a tunneling experiment where the trap is 
opened on one side and the out-going particle and its spin is detected \cite{gerhard2012} for
the case of $N_\uparrow=3$ and $N_\downarrow=1$.
{\bf B} Probability to find the spin down impurity on the far right for $N_\uparrow=1,\cdots,9$
in a $b=0$ harmonic trap (solid) and a hard wall trap (dashed) for the ground state. The dotted line is $1/N$ and
is the probability in both traps for the non-interacting state, $\Psi_\textrm{A}$. It is also the 
result if the $N_\uparrow$ majority particles are strongly interacting identical bosons.
The inset shows the same data on a double-log plot. The dash-dotted line is $1/N!$ for comparison.}
\label{polaron}
\end{figure*}

\vspace{1em}
\noindent {\bf Impurity problems.}
\noindent As another demonstration of
the nature of the strong coupling regime
we consider the case of a single spin down (impurity) interacting 
with a variable number of spin up fermions. An impressive
recent experiment has considered this system for $N\leq 6$
\cite{wenz2013}. As the energy at $1/g\to 0$ is degenerate
further insight into the strongly interacting regime has to come
from correlations in the $N$-body wave functions. Here we consider
the probability for the impurity to tunnel out of the trap
as shown in Fig.~\ref{polaron}A. In a simple model, we 
assume that due to the strong repulsion only the particle 
on the far right can tunnel out as the barrier is lowered (see Methods
for details). Since the 
wave function in (\ref{psi}) contains a superposition of states with
the impurity in different positions, the probability is simply
given by the amplitude for it to be on the right, $a_N$.
In Fig.~\ref{polaron}B we plot the impurity tunneling probability, $P_\downarrow=|a_N|^2$,
for $N=2,\ldots,10$ for the ground state of both a harmonic
trap and a hard wall potential (see Methods for details). 
The results for $N=2$ and $3$ are trap independent, while for $N>4$ 
we see a clear geometrical dependence. In particular, we find
that the scaling with $N$ is completely different; whereas 
by a fit we find approximately $P_\downarrow\propto N^{-3}$ for the 
hard wall, the
harmonic trap is not a power law but rather closer to 
a $1/N!$ behavior as seen in the inset of Fig.~\ref{polaron}B. The exact results thus allow
us to conclude that geometry has a strong effect on correlations in 
the wave function. The exact wave functions also show that the 
impurity has a peak in its probability density at the center of the trap. 
This is already true for $N=3$ and shows that for strongly 
interacting systems, studying the few-body limit gives
insight into the behavior of larger systems. Also note
that the combination of McGuire's solution (using periodic
boundary conditions) \cite{mcguire1965} and the local 
density in the trap \cite{astra2013} can only capture the 
energy in the strongly interacting regime but does not
reproduce the energy slope to order $1/g$.
The $1/N$ line in Fig.~\ref{2plus1}B applies to
the non-interacting state $\Psi_\textrm{A}$ (which is often referred to 
as the 'fermionized' state). 
More importantly, $1/N$ is also the probability
obtained if the spin up particles had instead been strongly interacting
bosons (see Methods for further details). This demonstrates
a very strong deviation from the common perception of the similarity of 
strongly interacting fermions and bosons in 1D.

\vspace{1em}
\noindent {\bf Discussion}\\
\noindent The strongly interacting regime is very difficult to access both
numerically and analytically due to the large degeneracies. 
The present approach finds the exact solution to linear order
in $1/g$ in a manner that automatically yields eigenstates that 
are adiabatically connected to the eigenstates for smaller values
of $g$. Combining our analytical approach with numerical techniques
that perform well in the weakly and intermediate strength regime
will allow us to access the quantitative and qualitative behaviour
of 1D systems in arbitrary confining geometries. Furthermore, our
approach provides a necessary starting point for including higher
orders in $1/g$\cite{busch1998,sen1999} and it represents an essential benchmark for 
numerical calculations. References \cite{gharashi2013,lindgren2013,deu2013}
contain recent numerical calculations very close to the $1/g=0$ 
limit. The results presented here are in agreement with those 
calculations. A leading numerical method used for 1D systems
is the density matrix renormalization group (DMRG) technique. 
As this is intrinsically a variational approach, it is not
clear how well it will perform in strongly interacting limit
for multi-component systems where the degeneracy of the spectrum
is large. Again our method can provide a fairly simple benchmark of DMRG
by using the exact slopes of the energy as $1/g\to 0$.

It is extremely important to note that with no extra effort 
our method obeys the Lieb-Mattis theorem
\cite{lieb1962} which states that the energies may
be ordered according to total spin, $S$, with the ground
state having the minimal $S$. It is tempting to conclude
that the exact solution can be obtained by constructing 
eigenstates with well-defined total spin at $1/g=0$, and 
many previous attempts to solve this problem have 
been based on spin algebra and spin mappings in one 
way or another.\cite{deuret2008,guan2009} However, this
construction is not unique and our method demonstrates that the condition 
is insufficient to determine the eigenstates 
to order $1/g$ for $N>3$ for multi-component systems
where $M>3$. Insisting on states that are eigenstates
of the total spin provides just one extra constraint
and this is not enough to determine all the coefficients 
$a_k$. A more direct way to see that the spin algebra 
approach is incomplete, is to note that the construction 
of the spin functions ignores the confinement $V(x)$.\cite{guan2009}
While it does yield eigenstates in the strict limit $1/g=0$, 
the spin states obtained for $N>3$
are generally not adiabatically connected to states at 
large but finite $g$. The method presented here overcomes
this naturally by extremizing the slope of the energy 
as the criteria that determines the eigenstates.

Several previous papers have introduced Bose-Fermi
and Fermi-Fermi duality mappings for interacting 1D 
systems.\cite{cheon1999,girar2004,girar2010,girar2011} This is a very
nice mathematical tool for transforming between 
strongly and weakly interacting systems, but it must
be stressed that these techniques cannot be used to 
solve the problem considered here. 
Many of these techniques start from the 
antisymmetrized non-interacting state and then 
multiplies by a factor that ensures that under
exchange of two particles the sign comes out 
correctly (plus for two bosons and minus for two fermions).
As we have clearly shown, the coefficients ($a_k$)
are generally {\it not} integer. 
For multi-component systems and for Bose-Fermi mixtures, 
duality mappings must be supplemented
by knowledge of the solution on either the fermion or 
boson side of the duality transformation.
Without our solution 
one would merely be mapping into another unsolved problem. 
On the other hand, combining our technique and
duality transformations we expand the class of solvable systems.

\vspace{1em}
\noindent {\bf Methods}\\
\noindent {\bf Details of the energy functional derivation.}
As noted in the main text, in the strongly interacting limit, $1/g\to 0$, 
the boundary conditions in (\ref{bound}) imply that the total 
wave function $\Psi$ must vanish whenever $x_i=x_j$. For two identical 
fermions, this is trivial since the Pauli principle dictates that 
both sides of (\ref{bound}) vanish. For non-identical particles, 
the wave function must still vanish when they overlap, but the
derivatives from each side can generally be different since the 
Pauli principle provides no restrictions. The basic idea of our method
is now very simple. First construct an antisymmetric function, $\Psi_\textrm{A}$, with
energy $E_\textrm{A}$ using single-particle states as described in the main text. 
The most general $N$-body wave function is shown in (\ref{psi}). 
The number of independent coefficients in (\ref{psi}) can be 
deduced from Pauli symmetry to be $M=N!/S$, where the symmetry factor is
calculated according to the number of groups of identical particles
in the system; $S=N_1!N_2!\ldots N_n!$ if there are $n$ groups 
of identical particles with $N_1,N_2,\ldots,N_n$ in each group.
Now we can
construct an energy functional, $K(a_1,\ldots,a_M)$, such that 
$E=E_\textrm{A}-K/g$. Subsequently we vary $K$ with respect to $a_k$ and 
diagonalize the resulting linear system to obtain the exact
eigenstates and slopes of the energy to linear order in $1/g$.
Intuitively, the functional gives the slopes of the energy so 
that the ground state on the repulsive side ($g>0$) 
around $1/g=0$ will maximize $K$, the 
first excited state will be the next extreme point, and so on.

The proof of (\ref{functional}) is an exercise in application of techniques 
from standard quantum mechanics. Using either perturbation theory or the 
Hellmann-Feynman theorem, we have 
\begin{eqnarray}\label{de}
&K=-\frac{\partial E}{\partial g^{-1}}=g^2\frac{\partial E}{\partial g}=&\nonumber\\
&\lim_{g\to\infty}\,g^2\frac{\sum_{i>j}\int\prod_{k=1}^{N}dx_k  |\Psi|^2\delta(x_i-x_j)}{\langle \Psi|\Psi\rangle},&
\end{eqnarray}
where the Dirac bracket $\langle \Psi|\Psi\rangle$ denotes the normalization 
integral. The dependence on $g$ can be eliminated by using (\ref{bound}) in (\ref{de}) which yields
\begin{widetext}
\begin{equation}
K=
\frac{1}{4}\frac{\sum_{i<j}^{}\int\prod_{i=1}^{N}\mathrm{d}x_i \delta(x_i-x_j)\left|\left
[\left(\frac{\partial}{\partial x_i}-\frac{\partial}{\partial x_j}\right)_{x_i-x_j\to0^+}-
\left(\frac{\partial}{\partial x_i}-\frac{\partial}{\partial x_j}\right)_{x_i-x_j\to0^-}\right]\Psi\right|^2 }{\langle\Psi|\Psi\rangle},
\label{slope}
\end{equation}
\end{widetext}
where it is important that one first evaluates the derivatives and then integrates out the delta function.
Note that if $i$ and $j$ are the indices of 
identical fermions, then the Pauli principle requires antisymmetry under permutation
and we get a vanishing contribution to $K$.
We can now split this integral into $N!$ sectors with different particle orderings, i.e.
\begin{widetext}
\begin{equation}
K=
\frac{1}{4}\frac{\sum_{i<j}^{}\sum_k\int_{\Gamma_k}\prod_{i=1}^{N}\mathrm{d}x_i \delta(x_i-x_j)\left|\left
[\left(\frac{\partial}{\partial x_i}-\frac{\partial}{\partial x_j}\right)_{x_i-x_j\to0^+}-
\left(\frac{\partial}{\partial x_i}-\frac{\partial}{\partial x_j}\right)_{x_i-x_j\to0^-}\right]\Psi\right|^2 }{\langle\Psi|\Psi\rangle},
\end{equation}
\end{widetext}
where we sum over permutation of the coordinates, $P_k$, and the integration regions, $\Gamma_k$, are such that
$x_{P_k(1)}<x_{P_k(2)}<\ldots<x_{P_k(N)}$.
This is very much in the spirit of the Bethe ansatz of course. In the Bethe ansatz, the assumption is that the two-body 
potential scatters without diffraction \cite{sutherland2004}. In our case, the very notion of scattering 
is compromised by the presence of the external trap and not even asymptotically can we talk about 
free particles in a general confining 1D geometry. Here we use instead the local properties
of the two-body interaction.

From every boundary where two
particles coincide we will obtain a factor $(a_k-a_p)^2$ times a derivative of $\Psi_\textrm{A}$. 
If they are identical fermions, then $a_k=a_p$ and the term vanishes, but to keep it 
general we do not make such an assumption here.
$K$ can thus be written as a sum of quadratic differences of the $a_k$ coefficients. Likewise, 
we may use the normalization 
$\langle\Psi|\Psi\rangle=\sum_k a_{k}^{2}$ (corresponding to unit normalization on each of the 
$M$ sectors in the expansion (\ref{psi})). Therefore $K$ can be written
\begin{equation}\label{matrix}
K=\frac{\sum_{k\neq p}(a_k-a_p)^2\alpha_{k,p}}{\sum a_k^2},
\end{equation}
where $k$ and $p$ run over the number of {\it independent} coefficients $M=N!/S$ and we 
have used the antisymmetry of $\Psi_\textrm{A}$ to eliminate the factor $1/4$. This is (\ref{functional})
of the main text.
The quantity $\alpha_{k,p}$ is defined as
\begin{equation}\label{coeff}
\alpha_{k,p}=\int_{\Gamma_k} \mathrm{d}x_1\ldots\mathrm{d}x_N 
\delta(x_i-x_j)\left|\left(\frac{\partial \Psi_\textrm{A}}{\partial x_i}\right)\right|^2.
\end{equation}
Again we first have to take the derivative before integrating over the delta function.
Here $P_k$ is a permutation of the coordinates which has the property that $x_i$ and $x_j$ are next to each other so
that they can interact while $p$ denotes a permutation, $P_p$, of the same kind but with $x_i$ and $x_j$ 
in reverse order. This shows why we do not need to put the index $p$ explicitly on the right-hand side since
it is uniquely specified for given $k$, $i$, and $j$.
Note that these integrals will generally also depend on the ordering of all the 
other particles besides $x_i$ and $x_j$ which is specified by $P_k$. 

The decisive observation is that the ground state in the vicinity of $1/g\to 0$ 
will be the state that maximizes the slope $K$.
In fact, all sets of $a_k$ that extremize $K$ define a wave function that is an eigenstate around $1/g\to 0$, and 
these will be orthogonal. This is proved as follows. First define a basis of states given by setting $a_k=1$ and 
$a_p=0$ for $p\neq k$, this defines a set of $M$ so-called bump functions
that all have energy $E_\textrm{A}$ through $\Psi_\textrm{A}$. We 
now apply degenerate perturbation theory to first order which yields a secular matrix (to be discussed below) 
whose eigenvalues are
the slopes $K_i$ and eigenstates are the correct eigenfunctions for $1/g\to 0$. The result now follows
from the linear variation method which states that the extremizing combinations are orthogonal eigenstates.
We have just shown that we can use either degenerate perturbation or variation to find the exact wave functions
for the ground state for $1/g\to 0$ and we obtain the slopes, $K$, automatically. 
It is straightforward to argue for the adiabatic connection of the ground state at $g=0$ and the ground state around $1/g\to 0^+$ 
for the lowest $E_\textrm{A}$ value possible where $\Psi_\textrm{A}$ constructed by one particle in each of the $N$ lowest
single-particle states. This follows from the Lieb-Mattis theorem and the fact that the 
largest total spin state is uniquely defined. For higher states one must be more careful 
in connecting the states and symmetry classifications at both weak and strong 
interactions is a very useful tool \cite{harshman2014}. However, we stress that 
symmetries (permutation group, parity invariance etc.) cannot be used to determine
the $a_k$ coefficients themselves for general external confinement and arbitrary
number of particles.

The determination of the amplitudes, $a_k$, now proceeds by linearization 
of the functional, $\partial K/\partial a_k=0$. This 
produces an eigenvalue equation of the form $A v =K v$, 
where $v$ is a vector with $a_k$ as entries, while 
$A$ is a symmetric matrix containing combinations of the $\alpha_{k,p}$
coefficients. By diagonalizing $A$ we obtain the 
orthogonal and complete set of eigenstates. This completes the 
proof of the solvability 
of the strongly interacting problem in an arbitrary 
confining potential in 1D to linear order in $1/g$.
The derivation above allows us to write down the effective
strong interaction Hamiltonian $H_\textrm{eff}=E_\textrm{A} I_M-(1/g)A$, where
$I_M$ is an $M$-dimensional identity matrix. $H_\textrm{eff}$ can 
be used to study additional perturbations on the system 
such as external electromagnetic fields in the 
strongly interacting regime. 

The exact solutions generically have different $a_k$ 
coefficients (in fact a subset of coefficients can 
even vanish in a given eigenstate).
This explains why it is very difficult to achieve 
convergent results in the strongly interacting limit 
using numerical techniques that are not optimized 
to take this into account. 
By taking different $\Psi_\textrm{A}$ with different energies
$E_\textrm{A}$ we can now build the entire spectrum which will 
consist of a ladder of states each with an 
$M$-fold degeneracy at $1/g=0$ and determine their slopes, $K$, 
around $1/g=0$. This is illustrated for ground state and
excited state manifolds in Fig.~\ref{2plus1} in the
main text.

Fermions with $I>2$ internal states or colors (such as an $SU(I)$ model) are
solved by exactly the same method but with a different $M$ depending on the 
number of such colors. A minor adjustment for strongly interacting bosons
is that when two identical bosons are interchanged in permutations $P_k$ 
and $P_p$ of (\ref{functional}), we must take
$a_k=-a_p$ to compensate the antisymmetry of $\Psi_\textrm{A}$ and consequently 
add a term $4\alpha_{kp}a_{k}^{2}$ to the numerator in (\ref{matrix}) to account for the interactions in (\ref{hamil}).
Mixtures of fermions and bosons run along the same lines. Our only assumptions
are that the particles have equal masses, the same interaction strength $g$ between 
all components,  and a confining potential, $V(x)$, 
which is the same for all particles.

\vspace{1em} 
\noindent {\bf Details for four-body systems.}
The two spin up and two spin down systems discussed in the text have the general 
wave function
\begin{equation}\label{2plus2eq}
\Psi=\left\{
  \begin{array}{l l}
     a_1 \Psi_\textrm{A} \; {\rm for} \; x_1<x_2<x_3<x_4 \;  (\uparrow \uparrow \downarrow \downarrow)  \\
    a_2 \Psi_\textrm{A}  \; {\rm for} \;  x_1<x_3<x_2<x_4 \; (\uparrow \downarrow \uparrow \downarrow) \\
    a_3 \Psi_\textrm{A}  \; {\rm for} \;  x_3<x_1<x_2<x_4 \; (\downarrow \uparrow \uparrow \downarrow) \\
    a_4 \Psi_\textrm{A}  \; {\rm for} \;  x_1<x_3<x_4<x_2 \; (\uparrow \downarrow \downarrow \uparrow) \\
    a_5 \Psi_\textrm{A}  \; {\rm for} \;  x_3<x_1<x_4<x_2 \; (\downarrow \uparrow \downarrow \uparrow) \\
    a_6 \Psi_\textrm{A}  \; {\rm for} \;  x_3<x_4<x_1<x_2 \; (\downarrow \downarrow \uparrow \uparrow) \\
      \end{array} \right.
\end{equation}
where we have fixed $x_1$ and $x_2$ to be spin projection up while $x_3$ and $x_4$ have spin projection down. 
The topology of each configuration is indicated by the arrows. Note that we have only written the 
independent pieces of the wave function, the remaining terms are dictated by the Pauli principle. 
In the example we consider the ground state manifold (lowest energy at $1/g=0$), 
meaning that $\Psi_\textrm{A}(x_1,x_2,x_3,x_4)$ is the antisymmetric function formed by occupying the four lowest
states in the confining potential. The functional for the energy around $1/g=0$ using this 
basis becomes
\begin{widetext}
\begin{equation}
K=\frac{\alpha_{1,2}(a_1-a_2)^2+\alpha_{2,3}(a_2-a_3)^2+\alpha_{2,4}(a_2-a_4)^2+\alpha_{3,5}(a_3-a_5)^2+\alpha_{4,5}(a_4-a_5)^2+\alpha_{5,6}(a_5-a_6)^2}{\sum_{k=1}^{6} a_{k}^{2}},
\end{equation}
\end{widetext}
where we notice the absence of a term with $\alpha_{3,4}$ since those configurations have no matching boundaries. By 
using parity invariance, one sees that $\alpha_{1,2}=\alpha_{5,6}\equiv\alpha$ and $\alpha_{2,3}=\alpha_{2,4}=\alpha_{3,5}=\alpha_{4,5}\equiv\beta$, so
that we have only two independent coefficients. By variation of $K$ with respect to $a_k$, we obtain the 
$A$ matrix which has the rather simple form
\begin{equation}
A=\left[\begin{matrix} 
\alpha & -\alpha & 0 & 0 & 0 &0 \\
-\alpha & \alpha+2\beta & -\beta & -\beta & 0 &0 \\
0 & -\beta & 2\beta & 0 & -\beta &0 \\ 
0 & -\beta & 0 & 2\beta & -\beta &0 \\ 
0 & 0 & -\beta & -\beta & \alpha+2\beta &-\alpha \\ 
0 & 0 & 0 & 0 & -\alpha &\alpha 
\end{matrix}\right],
\end{equation}
and acts on the vector $v=[a_1,a_2,a_3,a_4,a_5,a_6]^T$, where the superscript $T$ denotes the transpose.
Note how parity symmetry is explicitly featured in $A$ since it is symmetric when reading it from the top left corner to the bottom right corner 
or vice versa. The explicit forms of $\alpha$ and $\beta$ are
\begin{eqnarray}
&\alpha=\int_{x_1<x_2<x_4} \textrm{d}x_1\textrm{d}x_2\textrm{d}x_3\textrm{d}x_4 \delta(x_2-x_3) |\frac{\partial\Psi_\textrm{A}}{\partial x_3}|^2&\\
&\beta=\int_{x_1<x_3<x_2} \textrm{d}x_1\textrm{d}x_2\textrm{d}x_3\textrm{d}x_4 \delta(x_2-x_4) |\frac{\partial\Psi_\textrm{A}}{\partial x_4}|^2,&
\end{eqnarray}
where the difference between the two quantities is the ordering of coordinates in the integration region. Integrating out the delta function and renaming the variables in $\alpha$ ($x_4\to x_3$) shows that we have $x_1<x_2<x_3$ and $x_1<x_3<x_2$, respectively. An easy way
to think about the difference is that $\beta$ has the pair that interacts on the side of the system (left and right sides are equivalent due to parity invariance), while for $\alpha$ the interacting pair is in the middle.
These two are 
generally {\it not} the same. However, for the particular case of the hard wall confinement, it turns out that $\alpha=\beta$. In this 
case our results are consistent with the Bethe ansatz approach \cite{oelkers2006}. For the double 
well potential $\alpha\neq\beta$. This is perhaps a subtle point even to experts in the field and may explain a number of previous failed attempts to obtain the exact solution. By diagonalization of $A$ we obtain the slopes and wave functions which allows us to determine
the structures and probabilites discussed in the main text.

For the case where the two pairs of identical particles are instead bosons, there are additional 
interaction terms when two identical bosons are next to each other. The modification to the $A$ matrix is 
very simple and consists of adding a diagonal matrix $\textrm{diag}(4\beta,0,2\alpha,2\alpha,0,4\beta)$ to
$A$. The ground state is then of the simple form $v=[1,-1,1,1,-1,1]^T$, which is easily seen to 
correspond to $\Psi=|\Psi_\textrm{F}|$ where $\Psi_\textrm{F}$ is the wave function for spinless fermions, i.e. the 
Girardeau wave function \cite{girardeau1960}. Thus we get uniform probability of $1/3$ for each of 
the three configurations shown in Fig.~\ref{2plus1}. Again we notice the differences between 
strongly interacting bosons and fermions.

\vspace{1em} 
\noindent{\bf Details for polaron systems.}
The fermionic polaron system where a single spin down (often called an impurity for obvious reasons) 
interacts with a number of identical spin up fermions is handled in a similar fashion to 
the four-body system, although the $A$ matrix has an even simpler structure. 
Let the single spin down particle have coordinate $x_1$, and the $N-1$ spin up particles have coordinates
$x_2,\ldots,x_N$. There are $N!/(N-1)!=N$ independent $a_k$ coeffiecients. The wave function 
is now given by 
\begin{widetext}
\begin{equation}\label{polaroneq}
\Psi=\left\{
  \begin{array}{l l}
     a_1 \Psi_\textrm{A} \; {\rm for} \; x_1<x_2<x_3<\ldots<x_{N-1}<x_N \;  (\downarrow \uparrow \uparrow \ldots \uparrow\uparrow)  \\
    a_2 \Psi_\textrm{A}  \; {\rm for} \;  x_2<x_1<x_3<\ldots<x_{N-1}<x_N \; (\uparrow \downarrow \uparrow \ldots \uparrow\uparrow) \\
    a_3 \Psi_\textrm{A}  \; {\rm for} \;  x_2<x_3<x_1<\ldots<x_{N-1}<x_N \; (\uparrow \uparrow \downarrow \ldots \uparrow\uparrow) \\
		. \\
		. \\
    a_N \Psi_\textrm{A}  \; {\rm for} \;  x_2<x_3<\ldots<x_{N-1}<x_{N}<x_1 \; (\uparrow \uparrow \uparrow\ldots \uparrow\downarrow) \\
      \end{array} \right.
\end{equation}
\end{widetext}
It is now straightforward to obtain the slope which is given by 
\begin{equation}
K=\frac{\sum_{k=1}^{N-1}\alpha_{k,k+1}(a_k-a_{k+1})^2}{\sum_{k=1}^{N} a_{k}^{2}}.
\end{equation} 
As before we can use parity invariance of the confinement to conclude that $\alpha_{1,2}=\alpha_{N-1,N}$, $\alpha_{2,3}=\alpha_{N-2,N-1}$
and so forth. This means that the number of independent $\alpha_{i,j}$ is $N/2$ for $N$ even and $(N-1)/2$ for $N$ odd.
The corresponding $A$ matrix is tridiagonal with entries
\begin{widetext}
\begin{equation}\label{polmat}
A=\left[\begin{matrix} 
\alpha_{1,2} & -\alpha_{1,2} & 0 & 0 & \ldots & 0 & 0 &0 \\
-\alpha_{1,2} & \alpha_{1,2}+\alpha_{2,3} & -\alpha_{2,3} & 0& \ldots & 0 & 0 &0 \\
. & . & . & . & \ldots & . & . &. \\
. & . & . & . & \ldots & . & . &. \\
0 & 0 & 0 & 0 & \ldots & -\alpha_{2,3} & \alpha_{2,3}+\alpha_{1,2} & -\alpha_{1,2}   \\ 
0 & 0 & 0 & 0 & \ldots & 0 & -\alpha_{1,2} &\alpha_{1,2} 
\end{matrix}\right].
\end{equation}
\end{widetext}
The different coefficients of the matrix are 
\begin{equation}
\alpha_{k,k+1}=\int_{\Gamma} \prod_{l=1}^{N}\textrm{d}x_l \delta(x_{k+1}-x_1) |\frac{\partial\Psi_\textrm{A}}{\partial x_{k+1}}|^2,
\end{equation}
for $k=1,\ldots,N/2$ for $N$ even and $k=1,\ldots,(N-1)/2$ for $N$ odd. Here the integral is over the 
region $\Gamma={x_2<x_3<\ldots<x_1<x_{k+1}<\ldots<x_N}$.
Again we note that these constants are {\it not} 
equal for general potentials. However, once more the hard wall confinement is a truly special case. 
There one can prove that the $\alpha_{k,k+1}$ are equal and one recovers the Bethe ansatz results \cite{oelkers2006}. 

The results above can be applied to the case where $x_2,\ldots,x_N$ are identical bosons instead of 
fermions. For the Hamiltonian in (\ref{hamil}) with a single coupling, $g$, for all pair-wise interactions,
the ground state is the one found by Girardeau many years ago \cite{girardeau1960} since the Hamiltonian does not
distinguish between the identical bosons and the impurity. Alternatively, by counting interacting pairs
in each configuration one can show that in a hard wall confinement the two-component boson case
is obtained by adding a diagonal matrix to $A$ of the form $\textrm{diag}(2(N-2),2(N-3),2(N-3),\ldots,2(N-3),2(N-2))$.
Note that the inter- and intra-species interactions are the 
same for these two-component bosons and the coupling constant goes to infinity ($1/g\to 0$). This follows from the structure of our
Hamiltonian which does not distinguish between the two species. An interesting example of non-identical 
inter- and intra-species interactions can be found in Ref.~\onlinecite{zinner2013}

For general potentials with non-constant $\alpha_{k,k+1}$, the matrix to add will still be diagonal but
now the entries are sums over a set of $\alpha_{k,k+1}$ with a contribution from each adjacent pairs of identical bosons 
for a given configuration. Let us denote the two species as $A$ and $B$ and consider the example
with three $A$ type bosons and one $B$ type (acting as the impurity). In this case the 
four diagonal terms (corresponding to configurations $BAAA$, $ABAA$, $AABA$ and $AAAB$) are 
$2(\alpha_{2,3}$+$\alpha_{3,4})$ for $BAAA$ and also for $AAAB$, and $2\alpha_{1,2}$ for
$ABAA$ and also for $AABA$ (there is symmetry
around the center of the diagonal due to the parity symmetry of the potential).
Thus, in the limit $|g|\to\infty$, we have the 
ground state wave function $\Psi=|\Psi_\textrm{A}|$.
This has probability $1/N$ for the impurity to sit on the 
right-hand side of the system. This is a clear demonstration of the differences between fermions and bosons
when there are multiple internal states. Our formalism can also be applied to a mixture with $N_1$ single component
fermions and $N_2$ single component bosons. However, if the Bose-Fermi and Bose-Bose zero-range interactions
have different strength parameters, then the situation changes drastically and the ground state 
depends on the ratio of these strength parameters as they diverge to infinity \cite{zinner2013}.

\vspace{1em} 
\noindent{\bf Tunneling.}
The tunneling problem considered in the main text is considered within a simple model that can be 
justified in several different ways. Here we will use arguments based on 
the many-body tunneling theory of Bardeen \cite{bardeen1961}. It has recently
been shown \cite{rontani2012} that the Bardeen theory can reproduce experimental results on 
strongly interacting two-body systems \cite{gerhard2012} and we assume that this remains true
for more particles as studied in the main text. 

We assume that the exact eigenstate we have obtained is the 
initially stationary state that starts to decay as the barrier 
is lowered. This initial state can be written as a sum of eigenstates
of the new Hamiltonian with a lowered barrier, and we arrive at a 
dynamical problem. While this may sound very complicated, the new 
Hamiltonian is still assumed to be strongly interacting and particles
are still not allowed to exchange positions. The particle that was
initially closest to the barrier before it is lowered will therefore
all be the dominant contribution to the flux going out of the trap.
The initial state therefore contains the information about the tunneling
rates. This can be made quantitative by computing very 
accurate matrix elements for the rates. Since our main propose is to 
point out the difference of different initial states in tunneling 
experiments we postpone further tunneling calculations for future
studies.

\vspace{1em}
\noindent {\bf Acknowledgements}\vspace{0.5em}\\
We thank J. Arlt, D. Blume, G.~M. Bruun, F. Mila, K. Riisager, A. Svane, and S. Gammelmark for discussion
and feedback on the manuscript. 
This work was funded by the 
Danish Council for Independent Research DFF Natural Sciences and the 
DFF Sapere Aude program.\\

\vspace{1em}
\noindent {\bf Author contributions}\vspace{0.5em}\\
A.G.V. and N.T.Z. devised the project and developed the formalism in collaboration with 
M.V., A.S.J. and D.V.F. The numerical calculations were done by A.G.V. under the supervision
of D.V.F. and by N.T.Z. The initial draft of the manuscript was written by A.G.V., N.T.Z. and M.V. and all authors
contributed to the revisions that led to the final version.\\

\vspace{1em}
\noindent {\bf Competing financial interests}\vspace{0.5em}\\
The authors declare no competing financial interests.

\vspace{1em}
\noindent {\bf Corresponding author}\vspace{0.5em}\\
Correspondence to: N.~T. Zinner, zinner@phys.au.dk

\end{document}